\documentclass{evolang}
\usepackage{graphicx}
\usepackage{url}
\usepackage{authblk}
\usepackage{float}
\usepackage{xcolor, soul}

\usepackage{xr-hyper}

\usepackage{tikz}
\usetikzlibrary{positioning, shapes.geometric, arrows.meta, calc, decorations.pathreplacing, decorations.markings}

\tikzset{
    box/.style={draw, minimum height=1.2em, minimum width=2em, align=center},
    roundedbox/.style={draw, rounded corners, minimum width=2em, minimum height=1.5em, align=center},
    circlebox/.style={draw, circle, minimum size=1.2em, align=center},
    arrow/.style={-{Latex[width=2mm]}, thick},
    dottedarrow/.style={-{Latex[width=2mm]}, thick, dotted},
    embedding/.style={rectangle, draw, fill=red!20, minimum width=1.8em, minimum height=1.5em},
    projection/.style={rectangle, draw, fill=purple!20, minimum width=2em, minimum height=1.5em},
    batchnorm/.style={rectangle, draw, fill=brown!30, minimum width=2.2em, minimum height=1.5em},
    vgg/.style={trapezium, draw, trapezium angle=60, minimum height=1.6em, fill=yellow!40, shape border rotate=90,align=center},
    node distance=1.5cm and 1.2cm
}

\newcommand{\fref}[1]{Figure~\ref{#1}}
\newcommand{\tref}[1]{Table~\ref{#1}}
\newcommand{\eref}[1]{Equation~\ref{#1}}

\newcommand{\aref}[1]{Appendix~\ref{#1}}

\newcommand{\eg}{\emph{e.g.}}

\newcommand{\ie}{\emph{i.e.}}

\begin{document}

\title{Learning to Communicate Across Modalities: Perceptual Heterogeneity in Multi-Agent Systems}
\author[*1]{Naomi Pitzer}
\author[1]{Daniela Mihai}
\affil[*]{Corresponding Author: np5n22@soton.ac.uk}
\affil[1]{University of Southampton, United Kingdom}



\maketitle

\abstracts{Emergent communication offers insight into how agents develop shared structured representations, yet most research assumes homogeneous modalities or aligned representational spaces, overlooking the perceptual heterogeneity of real-world settings. We study a heterogeneous multi-step binary communication game where agents differ in modality and lack perceptual grounding. Despite perceptual misalignment, multimodal systems converge to class-consistent messages grounded in perceptual input. Unimodal systems communicate more efficiently, using fewer bits and achieving lower classification entropy, while multimodal agents require greater information exchange and exhibit higher uncertainty. Bit perturbation experiments provide strong evidence that meaning is encoded in a distributional rather than compositional manner, as each bit's contribution depends on its surrounding pattern. Finally, interoperability analyses show that systems trained in different perceptual worlds fail to directly communicate, but limited fine-tuning enables successful cross-system communication. This work positions emergent communication as a framework for studying how agents adapt and transfer representations across heterogeneous modalities, opening new directions for both theory and experimentation.}

\section{Introduction}
One research direction in language evolution seeks to explain how communication systems and structures can emerge from interaction among individuals without predefined linguistic conventions. Emergent Communication between artificial agents provides an experimental framework for investigating this process, allowing for observation into how agents develop and align communication systems through interaction under different environmental and cognitive constraints~\shortcite{ren2020compositionallanguagesemergeneural,chaabouni2020compositionalitygeneralizationemergentlanguages,lazaridou2018emergencelinguisticcommunicationreferential,guo2019emergencecompositionallanguagesnumeric,mihai2021learningdrawemergentcommunication,spiegel2025visualtheorymindenables}. In most existing models, agents are assumed to share the same perceptual representations of their environment, meaning that they effectively ground their protocols (\eg, symbols) in a common perceptual space~\cite{HARNAD1990335}. This overlooks an important dimension of communication: interlocutors can inhabit different perceptual worlds and still establish successful communication~\shortcite{HAGER2019717,Wang2021,evtimova2018emergentcommunicationmultimodalmultistep,comon}. In this work, we examine how perceptual misalignment between agents influences the structure and efficiency of emergent communication protocols. We show that such misalignment can lead to divergences in efficiency and consistency, and that agents can adapt their protocols to communicate successfully across different perceptual spaces.


\section{Experimental Setup}
\label{sec:game-overview}

\begin{figure}[h!]
\centering
\begin{tabular}{cc}
\hspace{-0.4cm}
\includegraphics[width=0.49\textwidth]{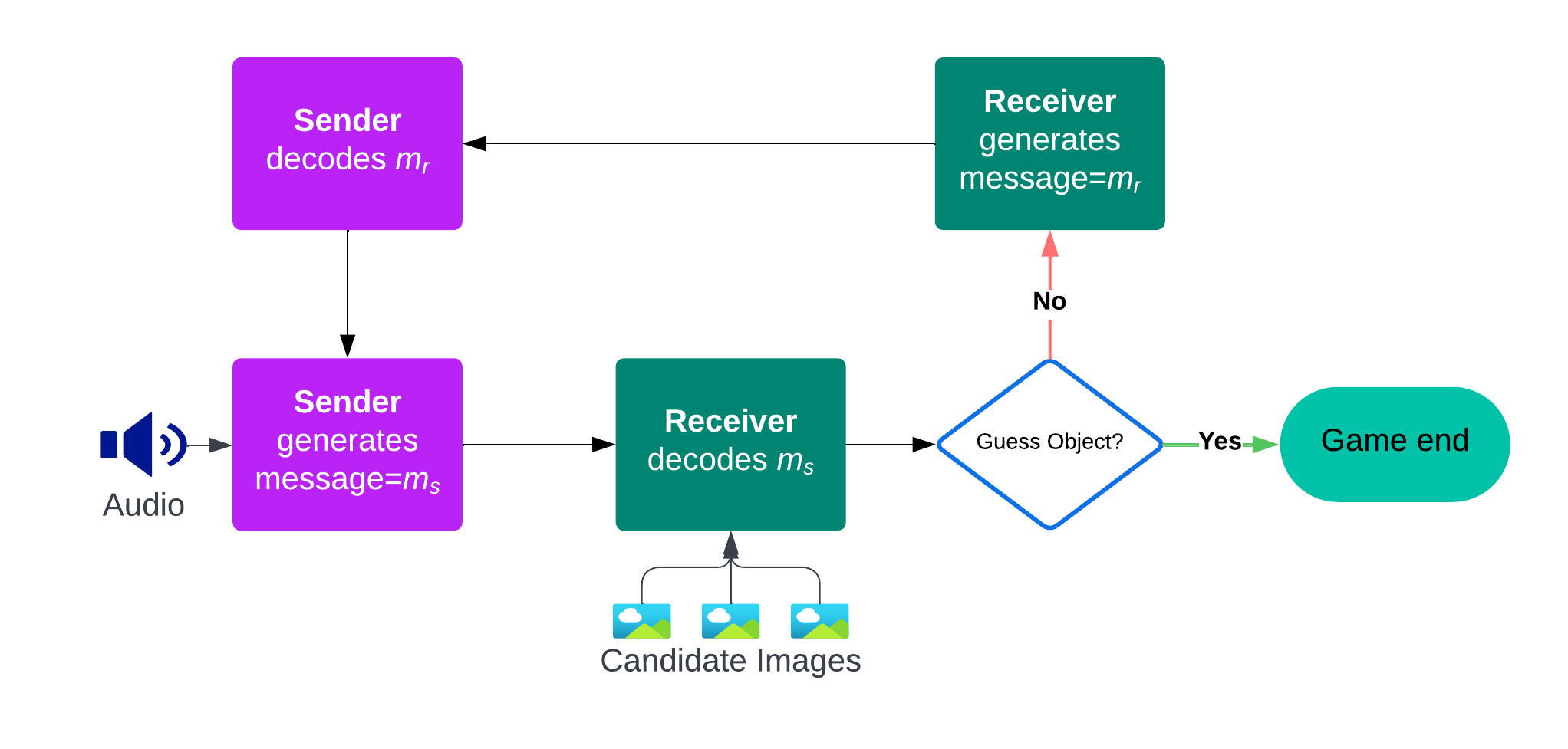} &
\includegraphics[width=0.49\textwidth]{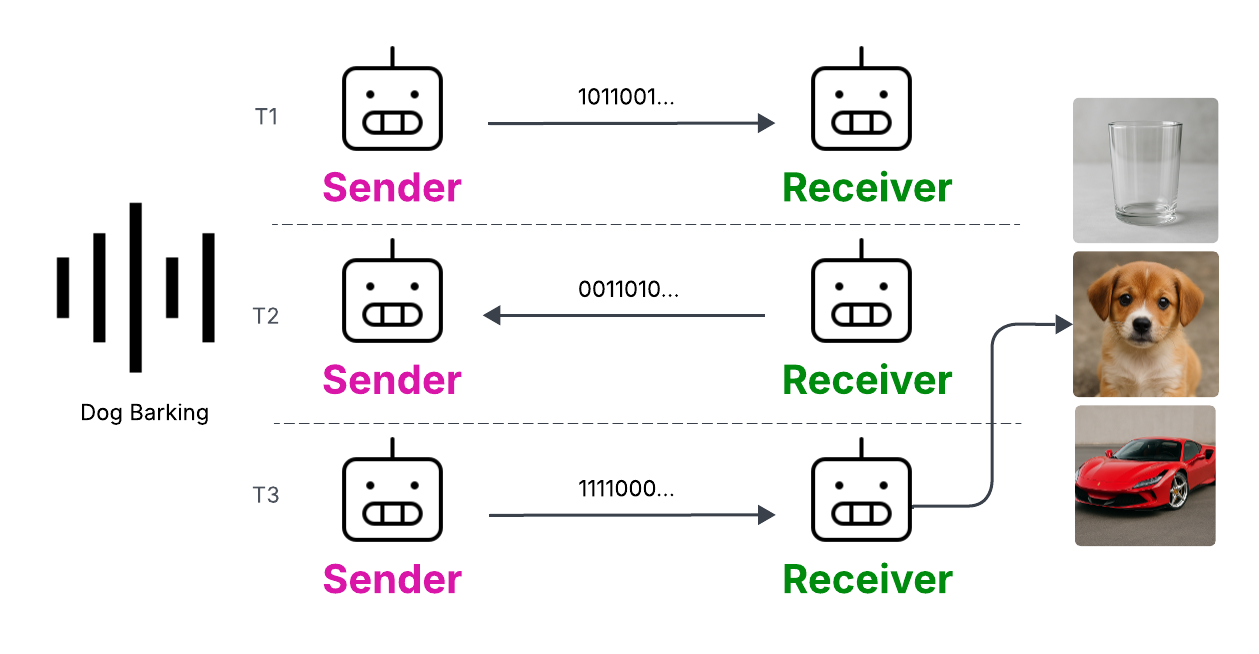}
\end{tabular}
\caption{{\footnotesize Illustration of the multi-step referential game, where a Sender encodes an audio input into a discrete message and a Receiver uses it to identify the correct visual target among a set of distractors, exchanging messages until a decision is made.}}
\label{fig:game-example}
\end{figure}

This work\footnote{
The code is available at https://github.com/naomipitzer/heterogeneous-emergent-communication. Further details in \aref{app:experimentalsetup}.} builds on the multimodal, multi-step referential game introduced by \shortciteA{evtimova2018emergentcommunicationmultimodalmultistep}, extending it from an image-text to an audio-image setting. In this game, a \textit{Sender} and \textit{Receiver} observe different modalities of the same target object (\eg, dog barking audio $\leftrightarrow$ image of a dog) and must communicate to identify the correct referent among a set of candidates (\fref{fig:game-example}). At each timestep $t$, the Sender emits a $D$-dimensional binary message ($m_t^s \in \{0,1\}^D$), which the Receiver uses to decide whether to terminate the dialogue ($s_t \in \{0,1\}$) and guess the object, or to respond with a message ($m_t^r \in \{0,1\}^D$). The exchange continues until $s_t = 1$ or the maximum step limit $T_{\max}$ is reached. To investigate the effects of heterogeneity, we also include a unimodal baseline (audio$\to$audio).

\vspace{-1em}
\vspace{-2em}

\paragraph{Agents.}
The Sender is a feed-forward neural network mapping its private input (\eg, audio clip) and the Receiver’s latest message to a binary vector (\fref{fig:senderdiagram}). The Receiver is a recurrent neural network that integrates the Sender’s messages and candidate embeddings to output a selected target, a stop signal, and a reply (\fref{fig:receiver-diagram}). Both agents use simple baseline networks for expected reward estimation (\fref{fig:baseline-Sender}). For detailed agent architectures, please refer to \aref{app:agents}.

\paragraph{Training and Data.}
The agents are jointly trained using RMSprop~\cite{tielman2012rmsprop} to minimise a combined objective integrating classification loss, REINFORCE-based reinforcement loss~\cite{williams1992}, and entropy regularisation. We use both a simple synthetic dataset called ``Shapes World" for controlled evaluation and a natural environmental dataset pairing CIFAR-100~\cite{cifar} images with UrbanSound8K~\shortcite{urban8k} and ESC-50~\cite{esc50} sounds to assess performance under realistic, noisy conditions.

\section{Results}
\label{sec:results}
We find that perceptual heterogeneity exerts a strong influence on the properties of the emergent language, namely on the efficiency, consistency, and interoperability of the learned protocols. 

\subsection{Perceptual Misalignment and Efficiency}
\label{sec:misali}

We first question whether perceptual heterogeneity, manifesting as perceptual misalignment, decreases communication efficiency. Here, efficiency is defined as the ability of the agents to compress internal representations into short, discrete messages \textit{while maintaining task accuracy}. To quantify this, message length is varied and the highest test accuracy achieved for both unimodal and multimodal systems is reported. Additionally, we measure the Receiver’s classification entropy to capture uncertainty in message decoding under increasing compression.

Both systems tolerate compression up to a threshold~(\tref{tab:msglength_supp}). However, when the message length is gradually reduced from 50 to 5 symbols, the systems' behaviour diverges. Unimodal agents maintain accuracy with low uncertainty, while \textbf{multimodal agents experience both degraded accuracy and increased uncertainty}. We interpret this divergence as a modality gap, where misaligned perceptual representations introduce noise into the communication channel. Such noise lowers the effective information transmitted per symbol, requiring greater capacity to sustain reliable communication~\shortcite{bottlenecktheory,sucholutsky2023alignmenthumanrepresentationssupports}. 
\begin{table}[h]
  \tablecaption{\footnotesize Effect of message length on communication efficiency and uncertainty.}
  {\begin{tabular}{lcccccccc}
\hline
\textbf{Length} &
\multicolumn{2}{c}{\textbf{Accuracy (\%)}} &
\multicolumn{2}{c}{\textbf{Loss}} &
\multicolumn{2}{c}{\textbf{Rec Entropy}} &
\multicolumn{2}{c}{\textbf{Sen Entropy}} \\
& M & U & M & U & M & U & M & U \\
\hline
1  & 34.00 & 36.00 & 1.82 & 1.81 & 1.19 & 1.03 & 16.95 & 14.98 \\
5  & 70.00 & 87.00 & 0.89 & 0.22 & 1.06 & 0.53 & 16.70 & 15.18 \\
10 & 81.00 & 82.00 & 0.64 & 0.47 & 0.68 & 0.55 & 20.75 & 20.77 \\
30 & 83.00 & 83.00 & 0.62 & 0.34 & 0.66 & 0.45 & 21.37 & 21.38 \\
50 & 85.00 & 87.00 & 0.53 & 0.36 & 0.67 & 0.48 & 21.48 & 21.62 \\
\hline
\end{tabular}}
\label{tab:msglength_supp}
\end{table}

\subsection{Class Consistency of Sender-Receiver Protocols}
\label{sec:consistency}
A central property of a stable communication system is the consistent association between signals and meanings across interlocutors~\shortcite{signallinggame,steels,lazaridou2017multiagentcooperationemergencenatural}. In this setting, assessing message consistency across instances of a class allows us to test whether Senders and Receivers rely on shared representations. Consistency within a class is measured as the average pairwise cosine similarity among messages belonging to that class.

\begin{figure}[H]
\vspace{-1.5em}
\centering
\begin{tabular}{cc}
\includegraphics[width=0.47\textwidth]{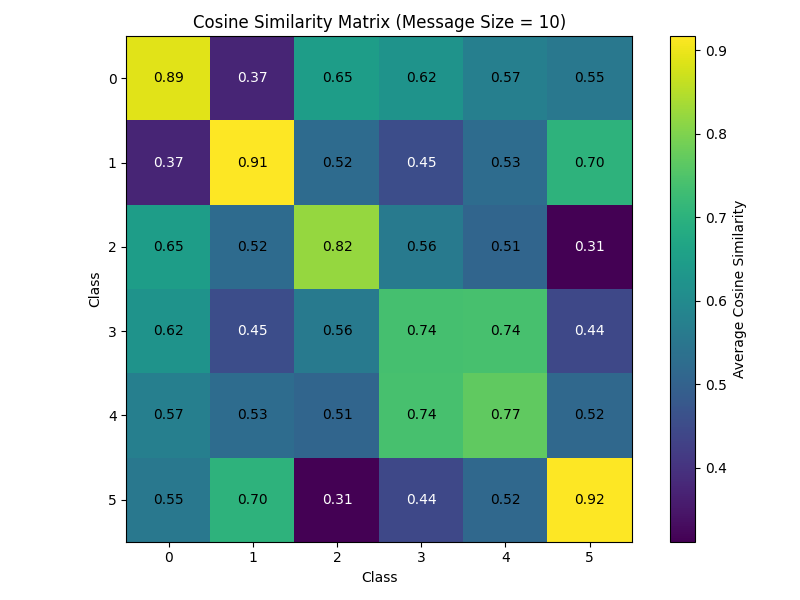} &
\includegraphics[width=0.47\textwidth]{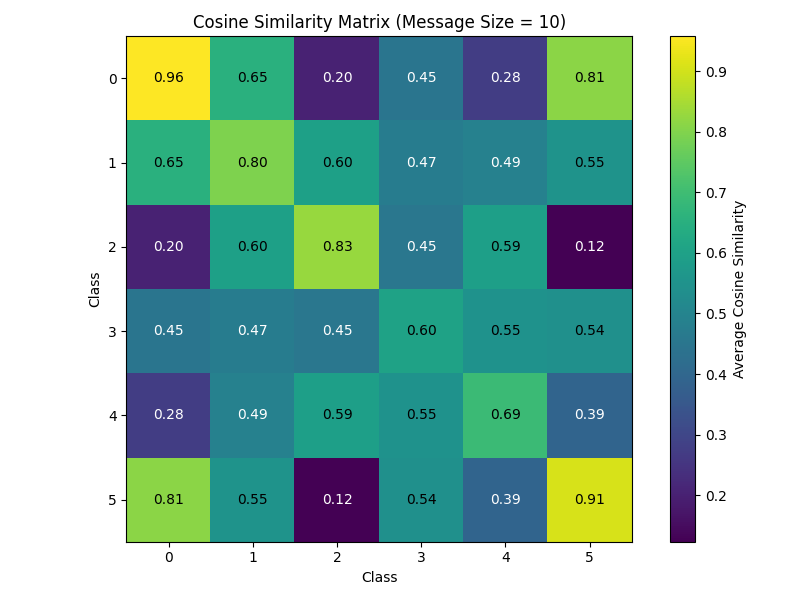} \\
\footnotesize{(a) Unimodal} & \footnotesize{(b) Multimodal}
\end{tabular}
\caption{{\footnotesize Cosine similarity matrices of Sender message embeddings for 
the unimodal and multimodal systems at message length 10. Sender messages within classes remain consistent in both unimodal and multimodal setups.}}
\label{fig:sendersimilarity_supp}
\end{figure}

\vspace{-.7em}
\paragraph{Sender/Receiver Internal Class Consistency}
Class-level consistency is measured separately for the Sender and Receiver to assess within-agent reliability. Sender messages exhibit strong class consistency, indicating convergence toward deterministic encodings (\fref{fig:sendersimilarity_supp}). Receiver messages, in contrast, show weaker internal consistency. This difference is expected: Receiver outputs depend not only on the target but also on the distractor set and class confidence, which vary across trials. Additionally, Sender consistency increases as message length decreases, suggesting that tighter capacity constraints encourage more deterministic protocols.

\vspace{-.7em}
\paragraph{Sender–Receiver Cross Class Consistency}
We compare Sender and Receiver messages per class to assess whether they converge to shared encodings. The results show little to no correlation exhibited between these messages, indicating that each agent develops its own protocol rather than relying on a common set of signals. 

\subsection{Grounding and Semantics}
\label{sec:grounding}
Having established that Sender and Receiver messages do not strongly correlate, we next examine how meaning is encoded within the Sender's messages and whether these encodings are grounded in perceptual input. All analyses are conducted on messages of fixed length 10. We use two approaches: perturbation experiments on message bits and message cluster analysis under varying perceptual features. 

\begin{figure}[H]
\centering
\begingroup
\definecolor{framegray}{gray}{0.7}%
\begin{tabular}{cc}
\includegraphics[width=0.45\textwidth]{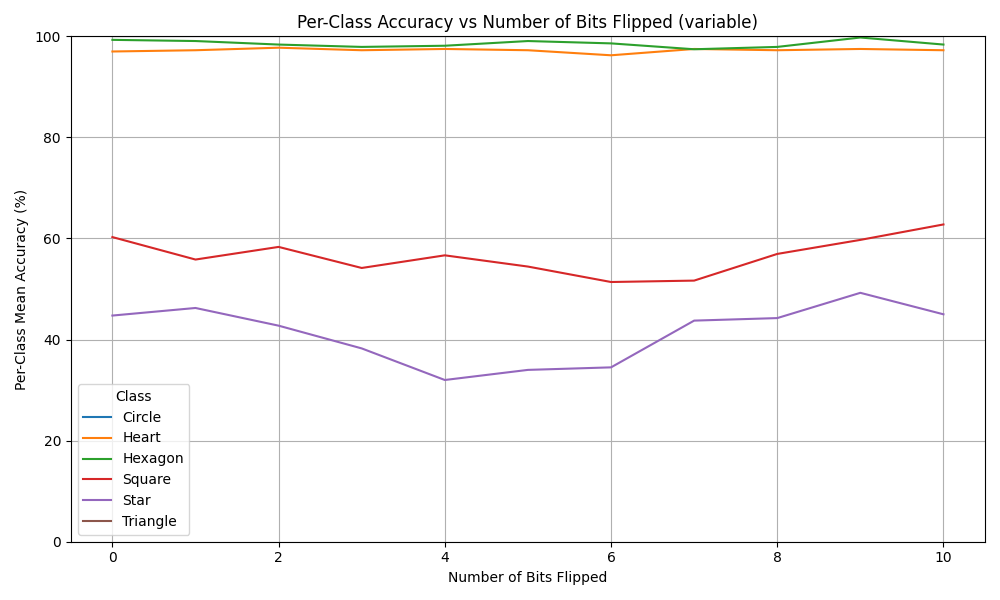} &
\includegraphics[width=0.45\textwidth]{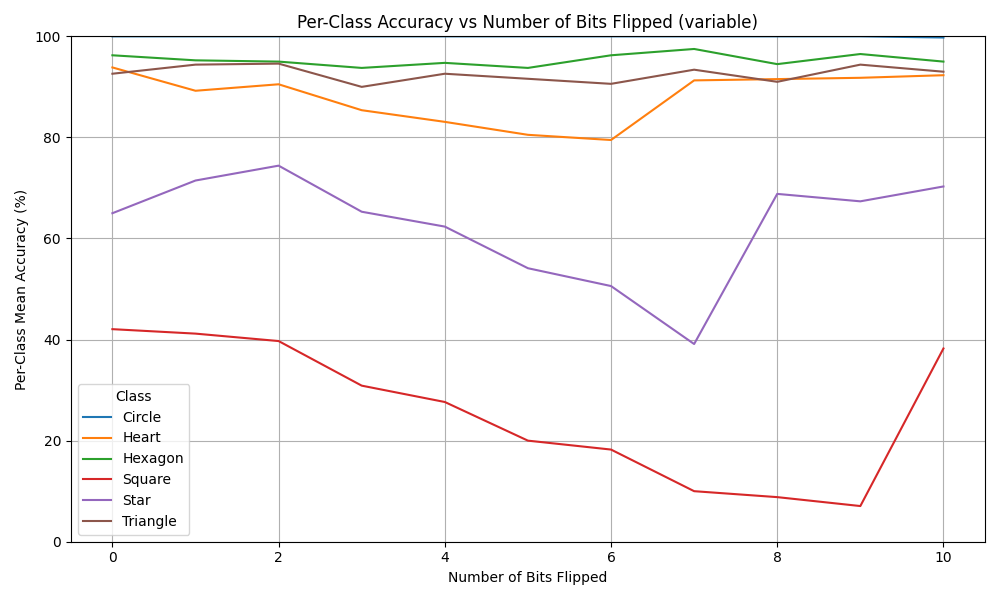} \\
{\footnotesize (a) Multimodal} & {\footnotesize (b) Unimodal}
\end{tabular}
\endgroup
\caption{{\footnotesize Per-class accuracy when flipping
variable (fluctuating) bits. Variable bit flips have minimal impact compared to constant bit flips, with a notably larger effect in the unimodal setup.}}
\label{fig:variablebitflipping}
\end{figure}

\vspace{-.7em}
\paragraph{Bit Perturbation.} Based on prior bit-usage analysis, bits are classified as \textit{constant} (\ie, those that are active (1) more than 90\% or inactive (0) less than 10\% of the time), or \textit{variable} (\ie, those that fluctuate between 0 and 1). Subsets of these bits are flipped, and the resulting classification accuracy and variance are measured. Conversations are restricted to a single exchange, since prior results show that the first message already conveys sufficient information for high task accuracy. These experiments reveal two key findings. Firstly, perturbing variable bits has minimal impact on accuracy in both unimodal and multimodal systems, while \textbf{flipping constant bits leads to sharp drops in accuracy}, indicating that they carry the most class-discerning information. However, perturbing variable bits in the unimodal case has a notably stronger impact on task accuracy than in the multimodal case, suggesting that these bits may encode the sender's fine-grained perceptual information that the multimodal Receiver, operating on a different modality, is less sensitive to (\fref{fig:variablebitflipping}).  

\begin{figure}[h!]
\centering
\begingroup
\definecolor{framegray}{gray}{0.7}%
\begin{tabular}{cc}
\includegraphics[width=0.45\textwidth]{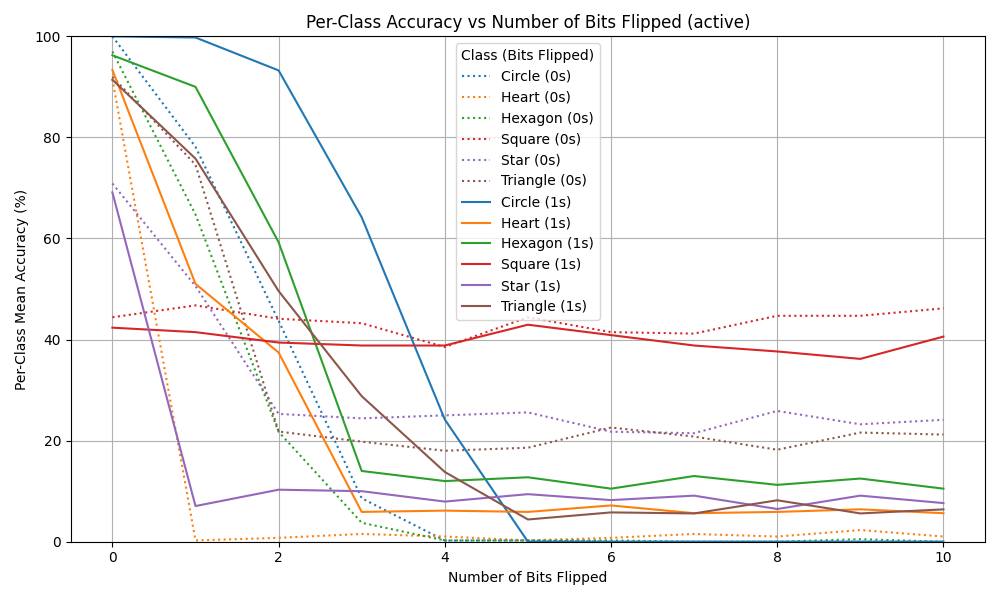} &
\includegraphics[width=0.45\textwidth]{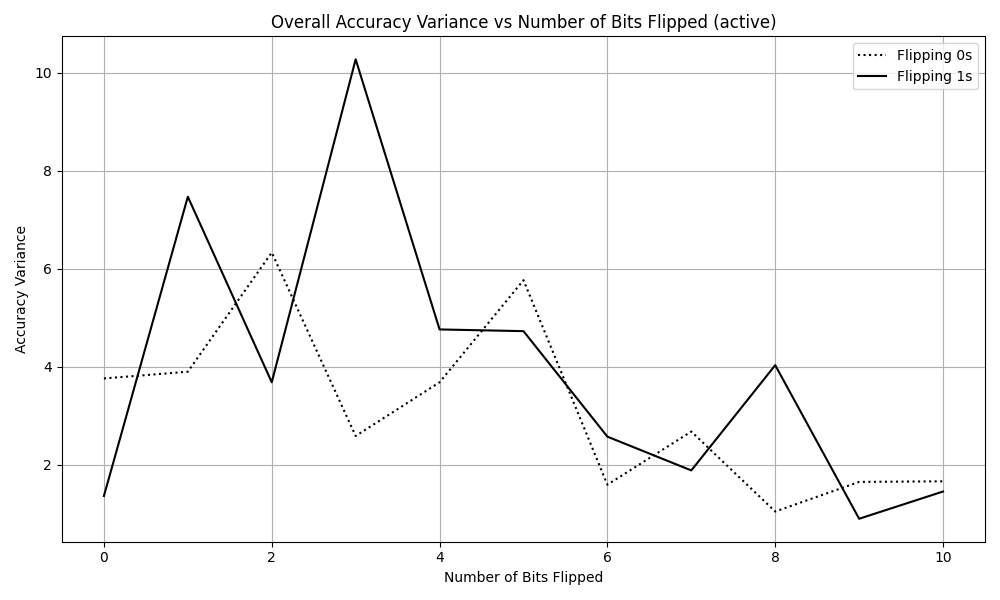} \\
{\footnotesize (a) Accuracy comparison} & {\footnotesize (b) Variance comparison}
\end{tabular}
\endgroup
\caption{{\footnotesize Per-class accuracy and variance when flipping an increasing number of constant bits: Flipping 0s to 1s generally leads to a steeper decline in accuracy than the reverse. The similar overall trends between flipping constant 0s and 1s suggest that both carry essential class information. A sharp spike in variance around the third bit flip when perturbing 1s indicates that certain always active bits encode disproportionately critical information.}}
\label{fig:constant-bit-analysis}
\end{figure}

In addition, certain constant bits carry disproportionate information, as reflected by sharp accuracy declines and increased variance when they are perturbed (\fref{fig:constant-bit-analysis}). However, these bits do not appear to encode information compositionally (\ie, their individual contributions cannot be combined to form interpretable meanings). For example, in the Heart class's case, flipping the final bit in the message from 0 to 1 reduced accuracy to the level of random guessing (approximately 16\%), yet the same bit was consistently inactive across other classes. This indicates that the bit cannot be assigned a fixed class-specific meaning. Instead, its \textbf{informational role is determined by the surrounding bit pattern}. 

\vspace{-0.7em}

\begin{figure}[H]
\centering
\begingroup
\definecolor{framegray}{gray}{0.7}%
\begin{tabular}{cc}
\includegraphics[width=0.45\textwidth]{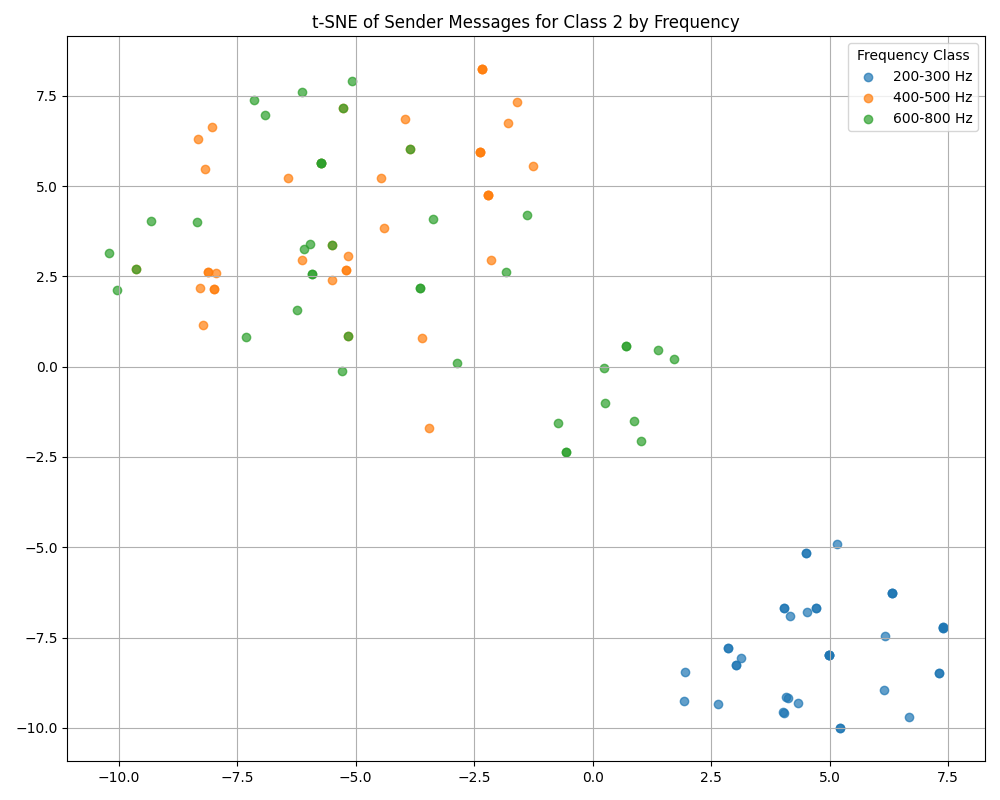} &
\includegraphics[width=0.45\textwidth]{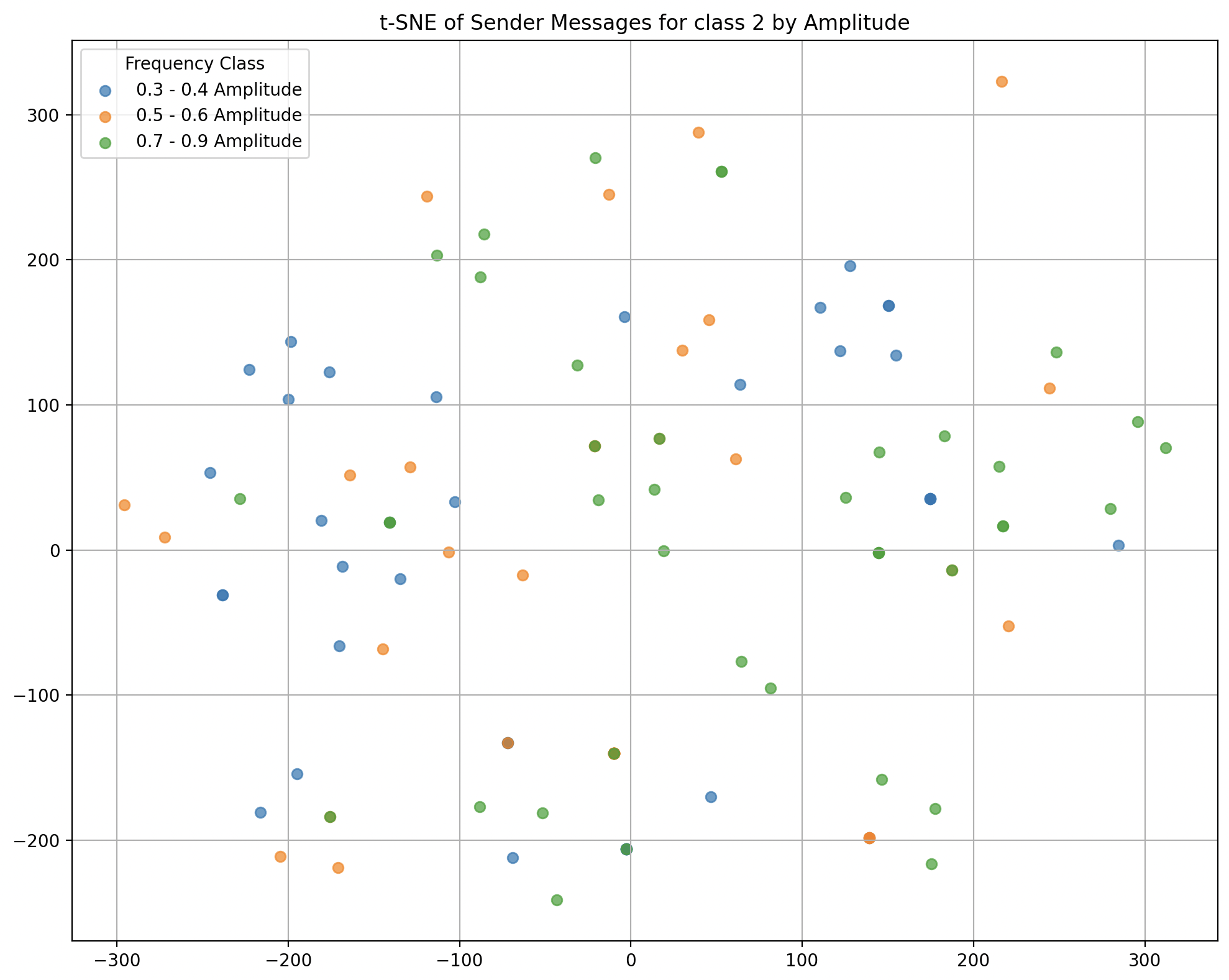} \\
{\footnotesize (a) Varying Frequency} & {\footnotesize (b) Varying Amplitude}
\end{tabular}
\endgroup
\caption{{\footnotesize t-SNE plots of Class 2 (heart class) messages across varying frequencies and
amplitudes. Messages corresponding to lower frequencies exhibit a consistent
underlying pattern, while variations in amplitude show minimal impact on message structure.}}
\label{fig:ampvsfreq_class2}
\end{figure}

\paragraph{Message Cluster Analysis.}
To further investigate perceptual grounding, controlled subsets of the synthetic audio dataset are constructed in which either audio frequency or amplitude varies while the other feature is held constant. We visualise the resulting message representations using t-SNE~\shortcite{tsne}. At the individual class level, frequency variation produces visible clusters: lower-frequency inputs form compact and separable groups (\fref{fig:ampvsfreq_class2}). 

\begin{figure}[h!]
\centering
\vspace{-1em}
\begin{tabular}{cc}
\includegraphics[width=0.49\textwidth]{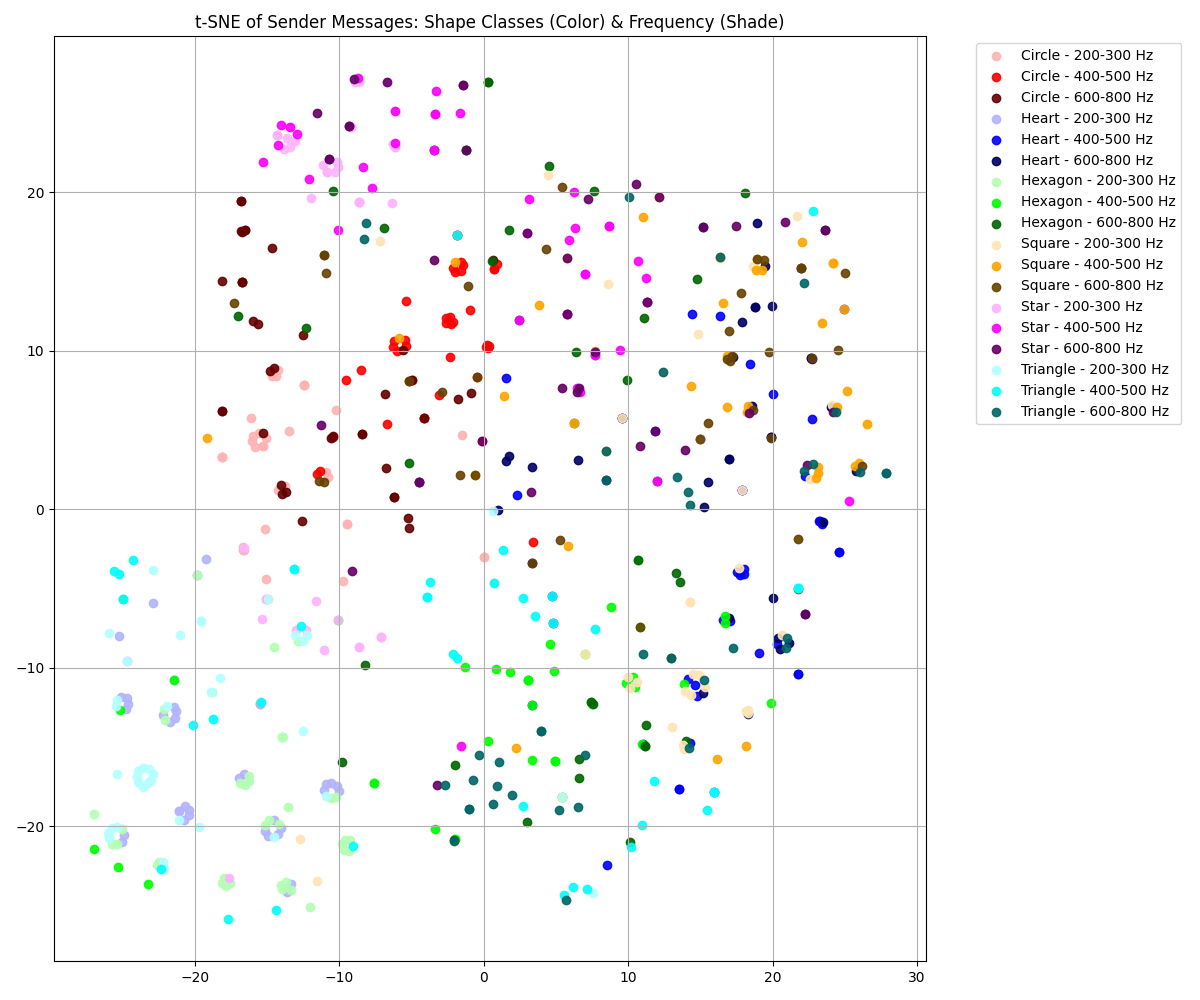} &
\includegraphics[width=0.49\textwidth]{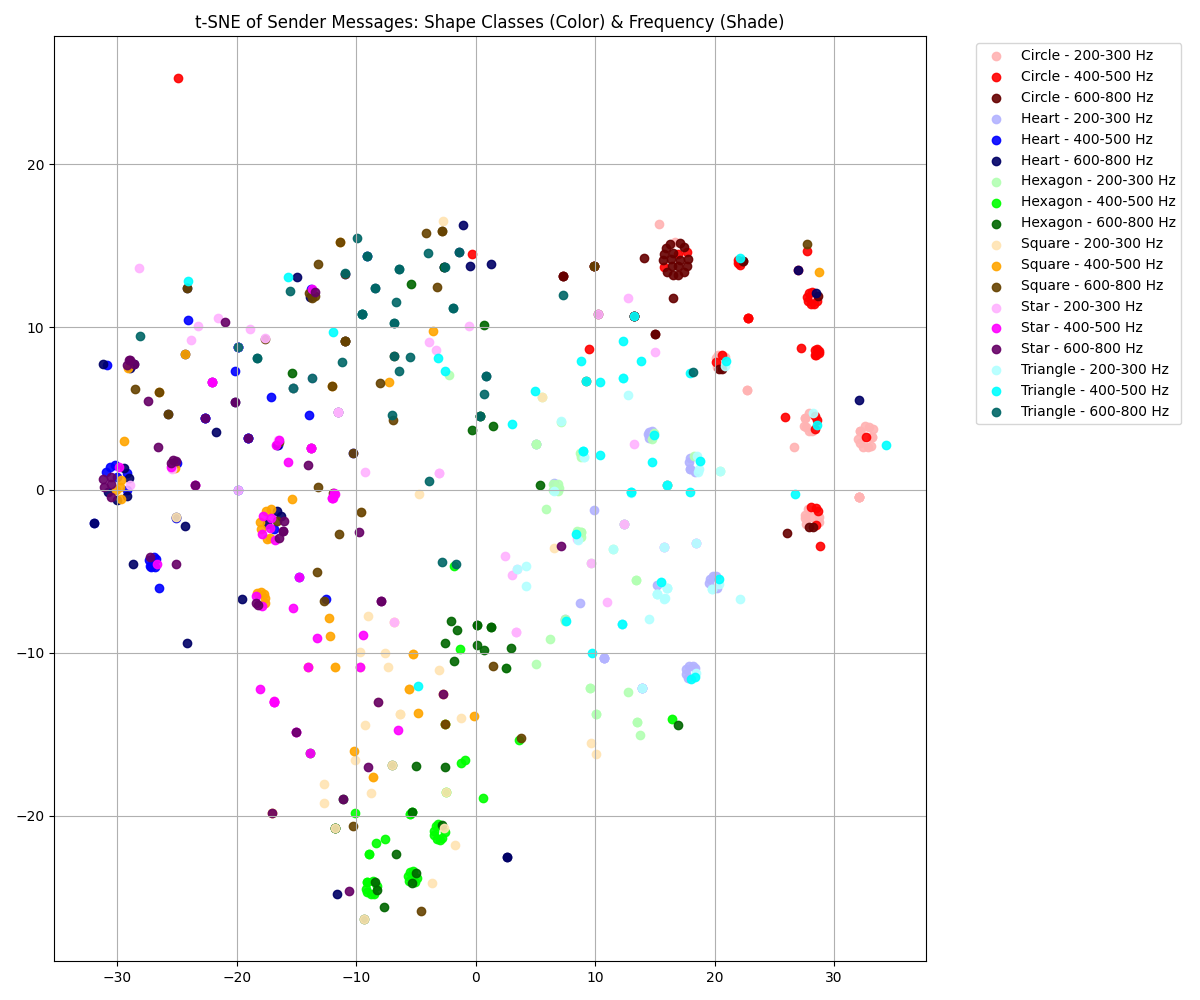} \\
{\footnotesize (a) Unimodal (audio–audio)} & {\footnotesize (b) Multimodal (audio–image)}
\end{tabular}
\caption{{\footnotesize t-SNE visualisations of \textbf{Sender message embeddings} for frequency variation in unimodal and multimodal systems.  
Panel (a) shows that in the unimodal setup, messages cluster strongly by frequency, indicating that low-level perceptual features are directly encoded in the communication protocol.  
Panel (b) shows the multimodal case, where clustering is less distinct but still systematic, suggesting that even when the Sender and the Receiver operate in different modalities, the message structure retains traces of grounded perceptual information related to frequency.}}
\label{fig:uni-vs-multi-freq}
\vspace{-.7em}
\end{figure}

Across the entire dataset, low-frequency inputs largely cluster in a shared region of the embedding space, indicating that frequency is encoded rather than randomly distributed (\fref{fig:uni-vs-multi-freq}). These effects are more pronounced in the unimodal system, yet remain visible in the multimodal system. These findings are notable, as they demonstrate that even when the Receiver perceives through a different modality, the Sender’s messages continue to reflect perceptual structure grounded in its modality. In other words, the Sender's messages in multimodal communication \textbf{remain grounded in the Sender’s perceptual space rather than becoming detached from it}. 

\subsection{Interoperability Across Modalities}
\label{sec:intero}
\begin{figure}[H]
\centering
\begingroup
\definecolor{framegray}{gray}{0.7}%
\begin{tabular}{cc}
\includegraphics[width=0.45\textwidth]{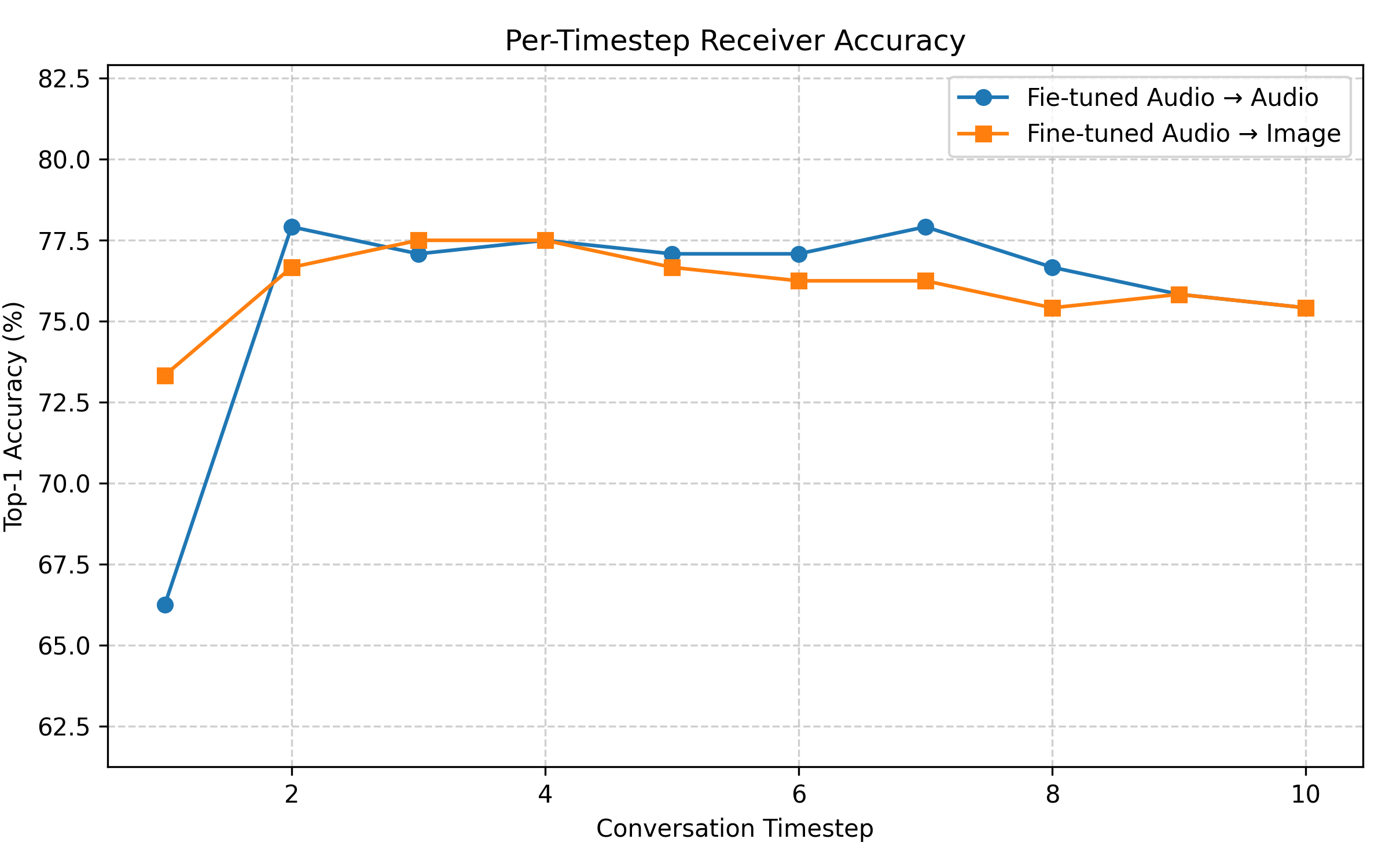} &
\includegraphics[width=0.45\textwidth]{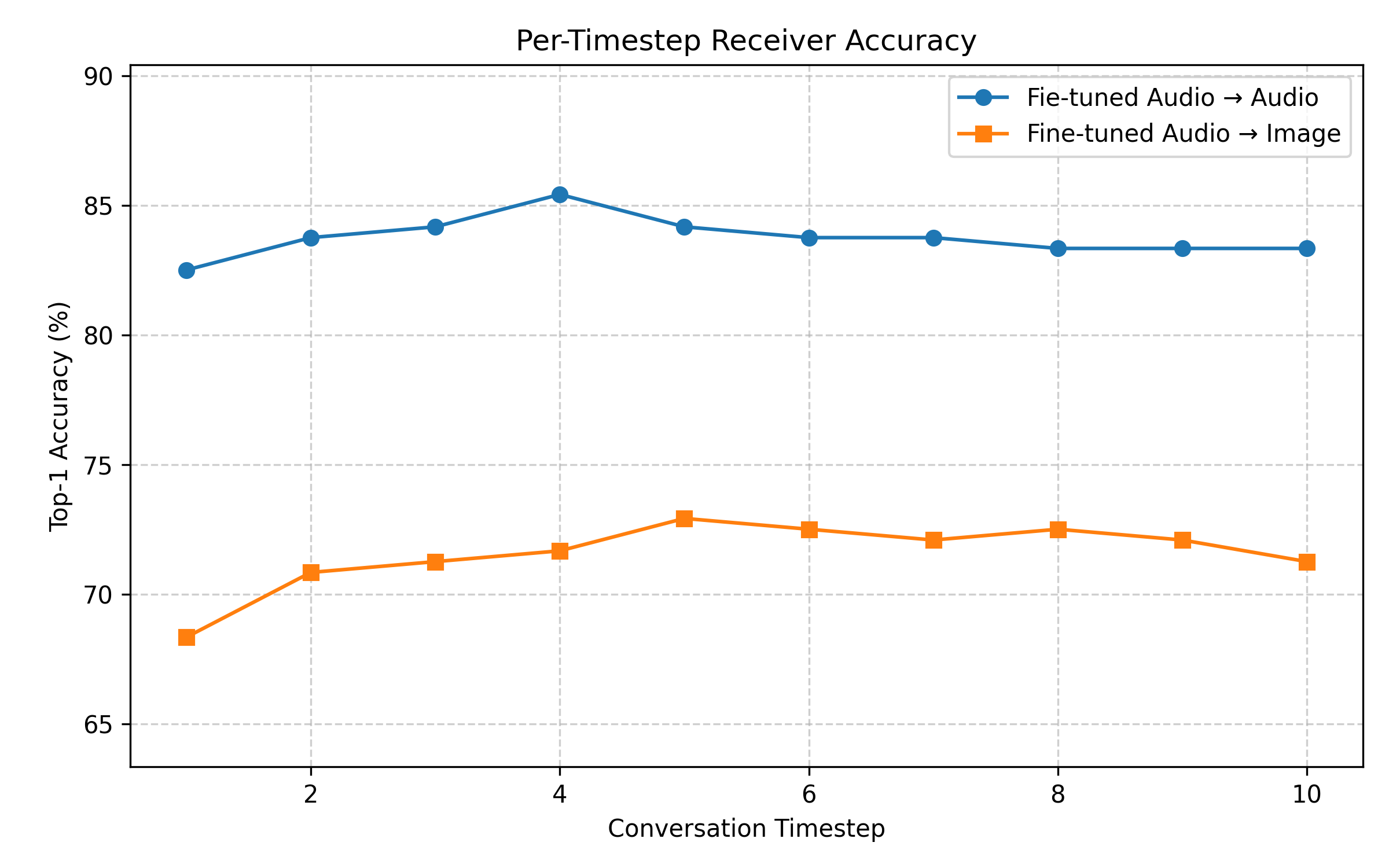} \\
{\footnotesize (a) 15 epochs} & {\footnotesize (b) 100 epochs}
\end{tabular}
\endgroup
\caption{{\footnotesize  System accuracy across conversation timesteps for a multimodal trained image-based and unimodal trained audio-based Receivers. The audio Receiver shows a sharp accuracy increase after the first exchange at 15 epochs of fine-tuning, while the multimodal Receiver remains stable. Both become consistent after extended fine-tuning.}}
\label{fig:interpoerability-pertimestep}
\end{figure}

We next ask whether agents trained in different contexts can communicate effectively with one another. Specifically, we pair an audio-trained multimodal Sender (originally trained with an image Receiver) with a unimodal audio Receiver. Without fine-tuning, accuracy falls to the level of random guessing (\tref{table:interoperability}), providing strong evidence that the emergent protocols are not directly interoperable across systems. We then apply various levels of fine-tuning to the paired system. Beyond 2 epochs, accuracy increases substantially, demonstrating that minimal adaptation suffices to align the Sender’s protocol with the unimodal Receiver. Extending fine-tuning further produces a trade-off, as accuracy on the newly paired system continues to improve, but at the cost of slightly degrading performance with the Sender’s original partner. At around 15 epochs of fine-tuning, both systems achieve high task accuracy simultaneously (\fref{fig:interpoerability-pertimestep}). Analysis of accuracy across conversation timesteps (Figure~\ref{fig:interpoerability-pertimestep}) reveals that after 15 epochs, the original multimodal system remains stable, with slight increases and decreases in accuracy throughout the conversation timesteps, while the audio-based system exhibits a sharp accuracy increase immediately after the first message exchange. This suggests that this first interaction might help the Sender identify its partner and adjust its communication protocol accordingly. After 100 epochs, both systems show stable communication throughout the exchange.

\begin{table}[ht]
  \tablecaption{Comparison of classification accuracy between an audio-based Receiver system and image-based Receiver system across different training durations. With minimal fine-tuning, the sender is able to establish communication protocols with different Receivers.}
  {\footnotesize
  \begin{tabular}{@{}ccc@{}}
  \hline
  \textbf{Number of Epochs} & \textbf{Audio Receiver Accuracy (\%)} & \textbf{Image Receiver Accuracy (\%)} \\
  \hline
  0 & 9.2 & 82.5 \\
  2 & 53.3 & 82.0 \\
  15 & 75.4 & 77.7 \\ 
  20 & 79.6 & 76.3 \\
  100 & 84.2 & 71.2 \\
  \hline
  \end{tabular}\label{table:interoperability}}
\end{table}

\section{Conclusions and Future Directions}
This study examined how perceptual input heterogeneity shapes the emergence, efficiency, and interoperability of communication protocols in artificial agents. The results show that when perceptual modalities are misaligned, the agents require longer messages and exhibit higher decoding uncertainty. Bit-level analyses further revealed that meaning is not localised to individual bits but distributed across the message pattern. Despite the input modality mismatch, the Sender's message embeddings remained grounded in its own perception, showing that communication can preserve perceptual structure even under representational divergence. Finally, we found that Senders can adapt their messages to communicate with Receivers experiencing differing perceptual realities. These findings highlight the asymmetries between unimodal and multimodal systems. Beyond the artificial settings, this work offers insight into how meaning and shared understanding can arise across divergent perceptual worlds, with implications for robotics, epistemology, and linguistic theories of embodiment and perceptual grounding. Future work should extend these analyses to alternative architectures, embodied agent systems (\eg, robots), and different embedding-generation strategies to develop a broader account of heterogeneous emergent communication.

\bibliographystyle{apacite}
\bibliography{evolang}
\clearpage
\appendix

\begin{center}
    {\LARGE\bfseries Appendices}
\end{center}

\addcontentsline{toc}{section}{Appendices}

\section{Detailed Experimental Setup}
\label{app:experimentalsetup}

This section provides the technical details of the experimental setup adopted for this work, which were briefly discussed in Section~\ref{sec:game-overview}.

\subsection{Agents}
\label{app:agents}

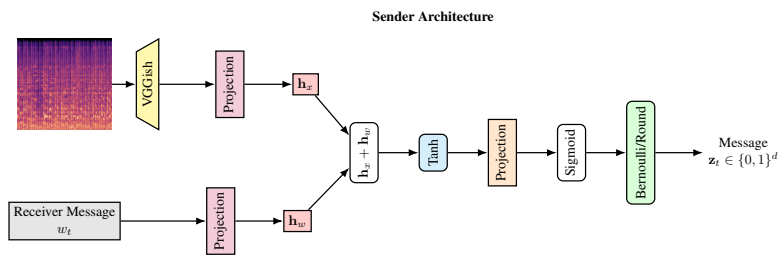
\begin{figure}[H]
  \centering
  \resizebox{0.9\linewidth}{!}{
  \begin{tikzpicture}[
      every node/.style={transform shape},
      node distance=1.8cm and 1.2cm,
      >=Latex,
      line width=0.9pt
  ]

  \node[anchor=south west, inner sep=0] (melimg)
      {\includegraphics[height=2.3cm]{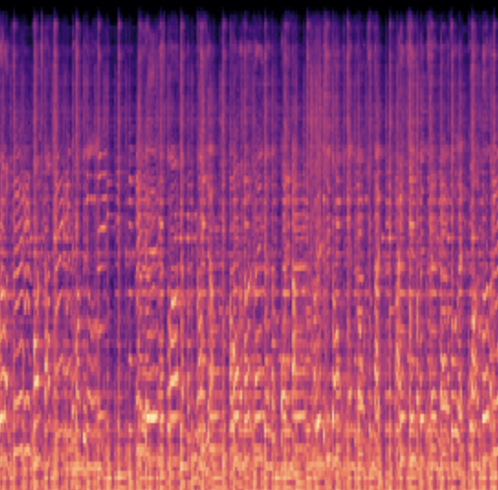}};

  \node[vgg, right=0.6cm of melimg, anchor=west] (vggs)
      {\rotatebox{90}{VGGish}};

  \node[draw, below=of melimg, fill=gray!20, minimum width=2.7cm, align=center] (recvmsg)
      {Receiver Message\\$w_t$};

  \draw[->] (melimg.east) -- (vggs.west);

  \node[projection, right=1.4cm of vggs, shape border rotate=90] (audio_proj)
      {\rotatebox{90}{Projection}};
  \node[projection, right=2.14cm of recvmsg, shape border rotate=90] (msg_proj)
      {\rotatebox{90}{Projection}};

  \node[embedding, right=of audio_proj] (hx) {$\mathbf{h}_x$};
  \node[embedding, right=of msg_proj] (hw) {$\mathbf{h}_w$};

  \path (hx) -- (hw) coordinate[midway] (mid);
  \node[roundedbox, right=of mid, shape border rotate=90] (sum)
      {\rotatebox{90}{$\mathbf{h}_x + \mathbf{h}_w$}};

  \node[roundedbox, right=1cm of sum, shape border rotate=90, fill=cyan!15] (tanh)
      {\rotatebox{90}{Tanh}};

  \node[projection, right=1cm of tanh, fill=orange!20, shape border rotate=90] (bin_proj)
      {\rotatebox{90}{Projection}};

  \node[roundedbox, right=1cm of bin_proj, shape border rotate=90] (sigmoid)
      {\rotatebox{90}{Sigmoid}};

  \node[roundedbox, right=1cm of sigmoid, fill=green!15, shape border rotate=90, align=center] (bernoulli)
      {\rotatebox{90}{Bernoulli/Round}};

  \node[right=of bernoulli, align=center] (output)
      {Message\\$\mathbf{z}_t \in \{0, 1\}^d$};

  \draw[->] (vggs) -- (audio_proj);
  \draw[->] (recvmsg) -- (msg_proj);
  \draw[->] (audio_proj) -- (hx);
  \draw[->] (msg_proj) -- (hw);
  \draw[->] (hx) -- (sum);
  \draw[->] (hw) -- (sum);
  \draw[->] (sum) -- (tanh);
  \draw[->] (tanh) -- (bin_proj);
  \draw[->] (bin_proj) -- (sigmoid);
  \draw[->] (sigmoid) -- (bernoulli);
  \draw[->] (bernoulli) -- (output);

  \node[above=2.7cm of tanh, align=center] {\textbf{Sender Architecture}};

  \end{tikzpicture}
  }

  \caption[\footnotesize Architecture of the Sender agent]{Architecture of the Sender agent (defined by \eref{eq:sender}). The Sender encodes audio using a VGGish model and combines it with an optional message using an element-wise sum of their linear projections to produce a communication signal. During training, the message is sampled from a Bernoulli distribution to allow gradient flow. This is replaced with a deterministic thresholding strategy during evaluation.}
  \label{fig:senderdiagram}
\end{figure}

\paragraph{Sender.}
The Sender is defined as:
\begin{equation}
A_S : O_S \times S \to S
\label{eq:sender}
\end{equation}
mapping its private input $o_s \in O_S$ and the Receiver's most recent message $m_{t-1}^r \in S$ to a binary message $m_t^s \in S$ where $S = \{0,1\}^D$ and $D$ is the length of the message. 
In our implementation, $o_s$ is an audio embedding extracted from a pretrained VGGish model, while in \shortciteA{evtimova2018emergentcommunicationmultimodalmultistep} it was an image embedding. 
The Sender combines both inputs through a linear projection of each, followed by an element-wise summation. The resulting fused representation is passed through a \texttt{tanh} activation layer and subsequently projected to the predefined message dimensionality. A final sigmoid layer constrains the output values to the $[0,1]$ range. During training, messages are generated by sampling each element from an independent Bernoulli distribution parameterised by these outputs, while at test time, they are thresholded deterministically. An overview of the Sender architecture is provided in Figure~\ref{fig:senderdiagram}.

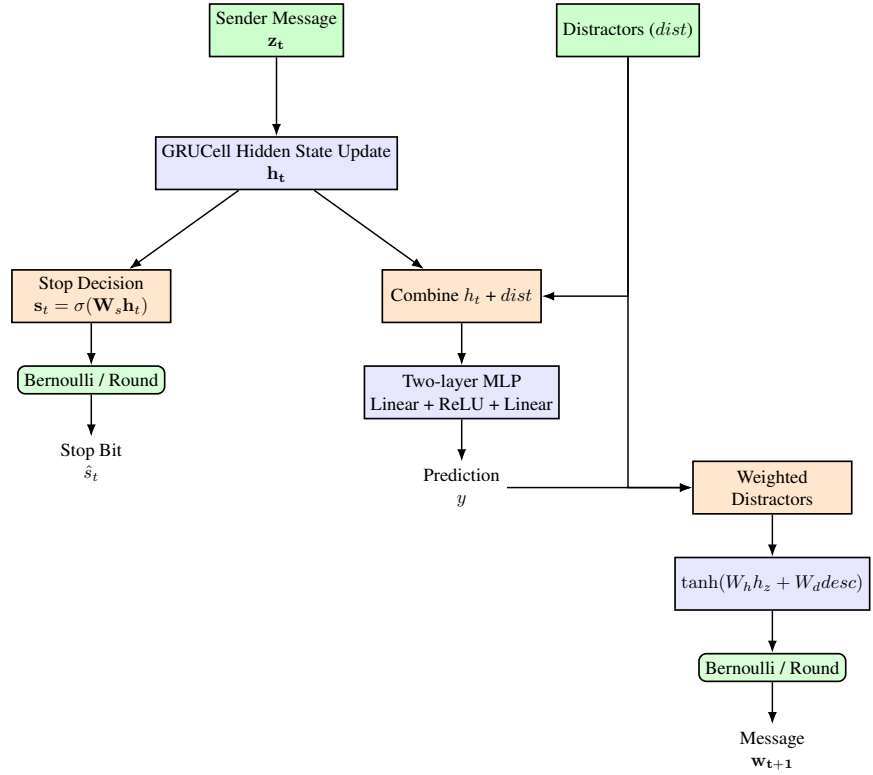
\begin{figure}[h!]
  \centering
  \resizebox{\linewidth}{!}{
  \begin{tikzpicture}[node distance=1.5cm, line width=0.9pt, >=Latex]

  \tikzstyle{layer} = [rectangle, minimum width=2.5cm, minimum height=1cm,
                       text centered, draw=black, fill=blue!10, align=center]
  \tikzstyle{input} = [rectangle, minimum width=2cm, minimum height=1cm,
                       text centered, draw=black, fill=green!20, align=center]
  \tikzstyle{process} = [rectangle, minimum width=3cm, minimum height=1cm,
                         text centered, draw=black, fill=orange!20, align=center]
  \tikzstyle{output} = [rectangle, minimum width=2.5cm, minimum height=1cm,
                        text centered, draw=black, fill=red!20, align=center]
  \tikzstyle{arrow} = [thick,->,>=Latex]

  \node (zinput) [input] {Sender Message \\ $\mathbf{z_t}$};
  \node (descinput) [input, right=4cm of zinput] {Distractors ($dist$)};

  \node (rnn) [layer, below=of zinput] {GRUCell Hidden State Update \\ $\mathbf{h_t}$};

  \node (stop) [process, below=of rnn, xshift=-3.5cm, align=center] {Stop Decision \\ $\mathbf{s}_t = \sigma(\mathbf{W}_s \mathbf{h}_t)$};
  \node[roundedbox, below=0.8cm of stop, fill=green!15, align=center] (bernoulli) {Bernoulli / Round};
  \node (stopout) [below=0.8cm of bernoulli, align=center] {Stop Bit \\ $\hat{s}_{t}$};

  \node (buildinp) [process, below=of rnn, xshift=3.5cm] {Combine $h_t$ + $dist$};
  \node (pred1) [layer, below=0.8cm of buildinp, align=center] {Two-layer MLP \\ Linear + ReLU + Linear};
  \node (yout) [below=0.8cm of pred1, align=center] {Prediction \\ $y$};

  \node (wd) [process, right=3.5cm of yout, align=center] {Weighted \\ Distractors};
  \node (commh) [layer, below=0.8cm of wd] {$\tanh(W_h h_z + W_d desc)$};
  \node[roundedbox, below=0.8cm of commh, fill=green!15, align=center] (bernoulli2) {Bernoulli / Round};
  \node (commout) [align=center, below=0.8cm of bernoulli2] {Message \\ $\mathbf{w_{t+1}}$};

  \draw [arrow] (zinput) -- (rnn);
  \draw [arrow] (rnn) -- (stop);
  \draw [arrow] (stop) -- (bernoulli);
  \draw [arrow] (bernoulli) -- (stopout);

  \draw [arrow] (rnn) -- (buildinp);
  \draw [arrow] (descinput) |- (buildinp);

  \draw [arrow] (buildinp) -- (pred1);
  \draw [arrow] (pred1) -- (yout);

  \draw [arrow] (yout) -- (wd);
  \draw [arrow] (descinput) |- (wd);

  \draw [arrow] (wd) -- (commh);
  \draw [arrow] (commh) -- (bernoulli2);
  \draw [arrow] (bernoulli2) -- (commout);

  \end{tikzpicture}
  } 

  \caption[Architecture of the Receiver network]{Abstracted architecture of the Receiver agent (defined by \eref{eq:receiver}). The Receiver takes as input the Sender's message and a set of distractors, updates its hidden state, computes stop decisions, generates predictions, and produces outgoing messages.}
  \label{fig:receiver-diagram}
\end{figure}

\paragraph{Receiver.}
The Receiver is defined as
\begin{equation}
A_R : S \times \mathbb{R}^q \to \Xi \times O_R \times S \times \mathbb{R}^q ,
\label{eq:receiver}
\end{equation}
where $\Xi = \{0,1\}$ is the stop decision, $O_R$ is the set of candidate objects, and $h_t \in \mathbb{R}^q$ represents the Receiver's internal hidden state at time step $t$.
Given the Sender’s message $m_s \in S$ and its previous hidden state $h_{t-1}$, the Receiver updates its state using a Gated Recurrent Unit (GRU), producing a new hidden state $h_t' \in \mathbb{R}^q$.
It then computes a probability distribution over all candidate objects $o_r \in O_R$, representing its current guess about the target object.
Finally, the Receiver decides whether to terminate the interaction ($s_t = 1$) or continue by generating a response message $m_r \in S$.
When $s_t = 1$, the Receiver selects the most likely image from the set of candidate images, $\hat{o}_r = \arg\max_{o_r \in O_R} p(o_r=1)$. In our setup, $O_R$ consists of either audio embeddings obtained from a VGGish network (unimodal system) or image embeddings obtained from a VGG16 network (multimodal system). Figure~\ref{fig:receiver-diagram} shows an abstracted illustration of the Receiver's architecture. 

\begin{figure}[h!]
  \centering
  \resizebox{\linewidth}{!}{
  \begin{tikzpicture}[
      every node/.style={transform shape},
      node distance=1.8cm and 1.2cm,
      >=Latex,
      line width=0.9pt
  ]

  \node[anchor=south west, inner sep=0] (melimg)
      {\includegraphics[height=2.3cm]{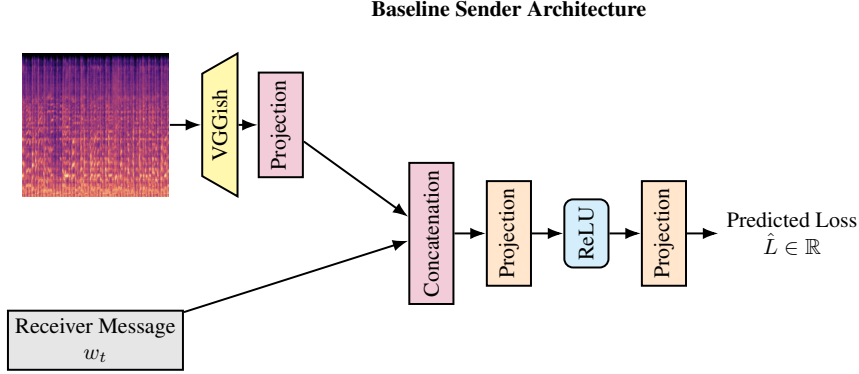}};

  \node[vgg, rotate=0, right=0.5cm of melimg, anchor=west] (vggs)
      {\rotatebox{90}{VGGish}};

  \node[projection, right=1.4cm of melimg, shape border rotate=90] (audio_proj1)
      {\rotatebox{90}{Projection}};

  \node[draw, below=of melimg, fill=gray!20, minimum width=2.7cm, align=center] (recvmsg)
      {Receiver Message\\$w_t$};

  \draw[->, line width=0.9pt] (melimg.east) -- (vggs.west);

  \path (vggs) -- (recvmsg) coordinate[midway] (vgmsgmid);
  \node[projection, right=4cm of vgmsgmid, shape border rotate=90] (audio_proj)
      {\rotatebox{90}{Concatenation}};

  \node[projection, right=0.5cm of audio_proj, fill=orange!20, shape border rotate=90] (bin_proj)
      {\rotatebox{90}{Projection}};

  \node[roundedbox, right=0.5cm of bin_proj, shape border rotate=90, fill=cyan!15] (relu)
      {\rotatebox{90}{ReLU}};

  \node[projection, right=0.5cm of relu, fill=orange!20, shape border rotate=90] (bin_proj2)
      {\rotatebox{90}{Projection}};

  \node[right=0.5cm of bin_proj2, align=center] (output)
      {Predicted Loss\\$\hat{L} \in \mathbb{R}$};

  \draw[->, line width=0.9pt] (audio_proj1) -- (audio_proj);
  \draw[->, line width=0.9pt] (recvmsg) -- (audio_proj);
  \draw[->, line width=0.9pt] (vggs) -- (audio_proj1);
  \draw[->, line width=0.9pt] (audio_proj) -- (bin_proj);
  \draw[->, line width=0.9pt] (bin_proj) -- (relu);
  \draw[->, line width=0.9pt] (relu) -- (bin_proj2);
  \draw[->, line width=0.9pt] (bin_proj2) -- (output);

  \node[above=2.5cm of bin_proj, align=center] {\textbf{Baseline Sender Architecture}};

  \end{tikzpicture}
  }

  \caption[\footnotesize Baseline Sender Architecture]{Architecture of the Sender Baseline agent. The Sender Baseline encodes the Sender's input using a VGGish model projected through the Sender's first hidden layer, and concatenates it with the current message. This feature vector is then passed through two Linear Layers to produce a scalar output representing the estimated loss.}
  \label{fig:baseline-Sender}
\end{figure}

\paragraph{Baseline.}
Both agents are equipped with baseline networks $B_S$ and $B_R$ that estimate expected rewards, reducing the variance of policy-gradient updates. 
These baselines are simple feedforward networks operating on the agent's current inputs and hidden state. $B_S$ receives as input the Receiver’s message and the Sender’s audio input projected through its first hidden layer (Figure~\ref{fig:baseline-Sender}), while $B_R$ takes as input the Receiver’s hidden state. Both networks output the predicted loss, $\hat{L} \in \mathbb{R}$.

\subsection{Training}
Agents are trained jointly for 200 epochs using RMSprop with a learning rate of $10^{-4}$. 
During each conversation, a new set of distractors is sampled and the conversation proceeds (\shortciteA{evtimova2018emergentcommunicationmultimodalmultistep} opted for a fixed distractor set) for up to $T_{\max}=10$ steps, with the Receiver free to terminate earlier. 
Variable-length conversations are handled with binary masks so that only active steps contribute to the loss. Gradient clipping is applied at every update for stability. The overall objective combines supervised classification, reinforcement learning, and entropy regularisation (\eref{eq:lossfunc}):
\begin{equation}
L_i = L^c_i + L^r_i - \sum_{t=1}^T \Big[ \epsilon_s H(s_t) + \epsilon_m \sum_{j=1}^d \big(H(m^s_{t,j}) + H(m^r_{t,j})\big) \Big]
\label{eq:lossfunc}
\end{equation}

where $H(\cdot)$ denotes entropy. The classification term is the negative log-likelihood (\eref{eq:nll}) of the correct candidate:
\begin{equation}
L^c_i = - \log p(o^*_r)
\label{eq:nll}
\end{equation}

The reinforcement learning term uses REINFORCE~\cite{williams1992} with baseline expected loss estimation subtraction (\eref{eq:reinforceloss}):
\begin{equation}
L^r_i = \sum_{t=1}^T (R - B_S) \sum_{j=1}^d \log p(m^s_{t,j}) 
      + \sum_{t=1}^T (R - B_R) \Big[\log p(s_t) + \sum_{j=1}^d \log p(m^r_{t,j})\Big]
\label{eq:reinforceloss}
\end{equation}
where $R$ is the reward given by the Receiver’s classification, and $B_S, B_R$ are baseline predictions for Sender and Receiver. 
The baseline networks are trained with mean squared error.

\subsection{Data Collection and Preprocessing}

\paragraph{Synthetic Dataset (Shapes World).}
To establish a controlled proof of concept, we construct a synthetic multimodal dataset consisting of six abstract classes: \textit{circle}, \textit{square}, \textit{triangle}, \textit{heart}, \textit{star}, and \textit{hexagon}. The dataset is designed to provide a structured and flexible environment for exploring emergent communication under noise-controlled and systematically variable conditions. A total of 2,400 images are generated using a Python-based image synthesis pipeline, with 400 samples per class. Each image depicts one of the six shapes rendered on a blank white background. Randomisation is applied to encourage generalisation while maintaining distinct inter-class boundaries. Specifically, for each instance, each shape is assigned a:
\begin{itemize}
\item \textbf{Colour:} random RGB, uniformly sampled from $[0, 255]$ for each channel.
\item \textbf{Shape Size:} dimensions vary between 50 and 150 pixels.
\item \textbf{Rotation:} random angle between $0^\circ$ and $360^\circ$ is applied to each shape.
\item \textbf{Position:} random centre coordinates such that the entire figure remains within the $256\times256$ pixel canvas.
\end{itemize}
In parallel, a corresponding audio dataset comprising 1,200 clips (200 per class) is generated using \texttt{NumPy} and \texttt{SciPy}. Each clip represents one of the six shape classes through a class-specific waveform pattern with controlled stochasticity. To simulate acoustic variation, several signal attributes are randomised:
\begin{itemize}
\item \textbf{Frequency:} sampled uniformly from 200Hz to 800Hz.
\item \textbf{Amplitude:} randomly selected from $[0.3, 0.9]$ (introduces loudness variation).
\item \textbf{Gaussian Noise:} Zero-mean Gaussian noise ($\sigma = 0.02$) is added.
\end{itemize}

For perturbation experiments, all other attributes are held constant while the frequency is separated into predefined ranges -- specifically, 200–300Hz, 400–500Hz, and 600–800Hz.

\paragraph{Environmental Dataset.}
To evaluate performance under more realistic and noisy conditions, we construct a multimodal environmental dataset by pairing natural images with environmental sounds. Six visual classes are selected from the CIFAR-100 dataset~\shortcite{cifar} (\textit{e.g.}, \textit{wolf}, \textit{lion}, \textit{pickup truck}), each containing 600 images resized and relabeled for consistency. Corresponding acoustic categories are drawn from the UrbanSound8K~\shortcite{urban8k} and ESC-50 datasets~\shortcite{esc50} (\textit{e.g.}, \textit{dog bark}, \textit{siren}, \textit{car horn}), also balanced to 600 samples per class. To address data scarcity, audio samples are augmented using the \texttt{audiomentations}~\shortcite{audiomentations} library with operations such as Gaussian noise addition, pitch shifting, time stretching, and gain adjustment. While some cross-modal mappings are imperfect (e.g., pairing \textit{wolf} images with ``dog bark" sounds), precise semantic alignment is not critical, as the goal is to examine how the system handles more complex data.

\paragraph{Audio Preprocessing.}
Every 0.96 seconds of audio inputs are transformed into 128-dimensional embeddings using a pretrained VGGish network~\shortcite{vggish}. These embeddings are then concatenated, and dimensionality is reduced to 128 dimensions using principal component analysis (PCA)~\shortcite{jolliffe2002pca}.

\paragraph{Image Preprocessing.}
Image inputs are embedded using VGG16~\shortcite{simonyan2015deepconvolutionalnetworkslargescale}. 
For the synthetic Shapes World dataset, the VGG16 model is fine-tuned on the generated shapes. 
For the environmental dataset, we use a pre-trained VGG16 without fine-tuning. The classification head is replaced with a 128-dimensional embedding layer, producing fixed-length feature vectors for each image. Embeddings are zero-mean normalised, as this was found to significantly improve the accuracy of the systems.

\end{document}